\documentclass[twocolumn]{aastex63}
\shorttitle{WASP-79b Eclipse}
\shortauthors{Foote et al.}
\graphicspath{{./}{figures/}}
\usepackage[version=4]{mhchem}
\usepackage{amsmath}

\begin{document}

\title{The Emission Spectrum of the Hot Jupiter WASP-79b from HST/WFC3}

\correspondingauthor{Trevor Foote}
\email{tof2@cornell.edu}

\author[0000-0002-6276-1361]{Trevor O. Foote}
\affiliation{Department of Astronomy and Carl Sagan Institute, Cornell University, 122 Sciences Drive, Ithaca, NY 14853, USA}

\author[0000-0002-8507-1304]{Nikole K. Lewis}
\affiliation{Department of Astronomy and Carl Sagan Institute, Cornell University, 122 Sciences Drive, Ithaca, NY 14853, USA}

\author[0000-0003-4220-600X]{Brian M. Kilpatrick}
\affiliation{Space Telescope Science Institute, 3700 San Martin Dr, Baltimore, MD 21218, USA}

\author[0000-0002-8515-7204]{Jayesh M. Goyal}
\affiliation{Department of Astronomy and Carl Sagan Institute, Cornell University, 122 Sciences Drive, Ithaca, NY 14853, USA}

\author[0000-0002-3288-0802]{Giovanni Bruno}
\affiliation{INAF - Catania Astrophysical Observatory, Via Santa Sofia, 78, 95123, Catania, Italy}

\author[0000-0003-4328-3867]{Hannah R. Wakeford}
\affiliation{School of Physics, University of Bristol, HH Wills Physics Laboratory, Tyndall Avenue, Bristol BS8 1TL, UK}

\author[0000-0002-8356-8712]{Nina Robbins-Blanch}
\affiliation{Department of Astronomy and Astrophysics, University of California, Santa Cruz, CA 95064, USA}

\author[0000-0003-3759-9080]{Tiffany Kataria}
\affiliation{Jet Propulsion Laboratory, California Institute of Technology, 4800 Oak Grove Drive, Pasadena, CA 91109, USA}

\author[0000-0003-4816-3469]{Ryan J. MacDonald}
\affiliation{Department of Astronomy and Carl Sagan Institute, Cornell University, 122 Sciences Drive, Ithaca, NY 14853, USA}

\author[0000-0003-3204-8183]{Mercedes L{\'o}pez-Morales}
\affiliation{Center for Astrophysics, Harvard \& Smithsonian, 60 Garden Street, Cambridge, MA 02138, USA}

\author[0000-0001-6050-7645]{David K. Sing}
\affiliation{Department of Earth \& Planetary Sciences, Johns Hopkins University, Baltimore, MD, USA}
\affiliation{Department of Physics \& Astronomy, Johns Hopkins University, Baltimore, MD, USA}

\author[0000-0001-5442-1300]{Thomas Mikal-Evans}
\affiliation{Department of Physics and Kavli Institute for Astrophysics and Space Research, Massachusetts Institute of Technology, Cambridge, USA}

\author[0000-0002-9148-034X]{Vincent Bourrier}
\affiliation{Observatoire Astronomique de l'Universit\'e de Gen\`eve, Chemin Pegasi 51b, CH-1290 Versoix, Switzerland}

\author[0000-0003-4155-8513]{Gregory Henry}
\affiliation{Center for Excellence in Information Systems, Tennessee State University, Nashville, TN 37209, USA}

\author[0000-0003-1605-5666]{Lars A. Buchhave}
\affiliation{DTU Space,  National Space Institute, Technical University of Denmark, Elektrovej 328, DK-2800 Kgs. Lyngby, Denmark}



\begin{abstract}

Here we present a thermal emission spectrum of WASP-79b, obtained via \textit{Hubble Space Telescope} Wide Field Camera 3 G141 observations as part of the PanCET program. 
As we did not observe the ingress or egress of WASP-79b's secondary eclipse, we consider two scenarios: a fixed mid-eclipse time based on the expected occurrence time and a mid-eclipse time as a free parameter. In both scenarios, we can measure thermal emission from WASP-79b from 1.1-1.7~$\mu$m at 2.4$\sigma$ confidence consistent with a 1900~K brightness temperature for the planet. We combine our observations with \textit{Spitzer} dayside photometry (3.6 and 4.5~$\mu$m) and compare these observations to a grid of atmospheric forward models that span a range of metallicities, carbon-to-oxygen ratios, and recirculation factors. Given the strength of the planetary emission and the precision of our measurements, we found a wide range of forward models to be consistent with our data.
The best match equilibrium model suggests WASP-79b's dayside has a solar metallicity and carbon-to-oxygen ratio, alongside a recirculation factor of 0.75. Models including significant H- opacity provide the best match to WASP-79b's emission spectrum near 1.58~$\mu$m. However, models featuring high-temperature cloud species - formed via vigorous vertical mixing and low sedimentation efficiencies - with little day-to-night energy transport also match WASP-79b's emission spectrum. Given the broad range of equilibrium chemistry, disequilibrium chemistry, and cloudy atmospheric models consistent with our observations of WASP-79b's dayside emission, further observations will be necessary to constrain WASP-79b's dayside atmospheric properties.

\end{abstract}



\vspace{5em}
\section{Introduction} \label{sec:intro}

Although our understanding of exoplanets and their atmospheres has greatly expanded in recent years, we still often lack the information necessary to probe the multi-dimensional nature of these worlds beyond our solar system. The combined insights from both spectroscopic transmission and emission observations have proved particularly useful in fleshing out the global thermochemical state of exoplanets (e.g. \citealt{Kreidberg_2014}). To date, the \textit{Hubble Space Telescope} (HST) has been the primary facility for spectroscopic emission observations for exoplanets that orbit close to their host stars. The near-infrared wavelength coverage offered by HST does present a challenge in probing the atmospheres of cooler exoplanets, but is ideal for measuring emission from hot Jupiters.

One such hot Jupiter, WASP-79b, is a target of particular interest for both transmission and emission observations. WASP-79b orbits ($a=0.0519$~AU) a relatively bright (V\textsubscript{mag} = 10) F-type star ($M_{\star}=1.43~M_\odot$, $T_{\star}=6600$~K), which makes it a target amenable to high-precision time-series observations with HST. 
The physical properties of WASP-79b ($R_p=1.53~R_J$, $M_p=0.86~M_J$, $T_{eq}\sim1700-1900$~K, \citealt{Brown_2017}) place it within transition regions of size and temperature phase space interesting for the study of cloud formation in the atmospheric regions probed via transmission observations \citep[e.g.][]{Stevenson_2016} and dayside atmospheric thermal inversions probed via eclipse observations \citep[e.g.][]{Baxter_2020}. Thus WASP-79b is a key target to better understand the processes shaping the thermochemical structure and cloud formation in hot Jupiters to further inform atmospheric modeling efforts \citep{Sudarsky_2003, Marley_2013}. In general, WASP-79b is a target of high importance to the exoplanet community, as highlighted by its status as one of the top candidate target for the \textit{James Webb Space Telescope} (JWST) Transiting Exoplanet Community Early Release Science (ERS) program (ERS Program 1366, PI Batalha, see \citealt{Stevenson_Lewis_2016,Bean_2018}).

Transmission spectroscopy of WASP-79b has revealed evidence of several chemical species at its day-night terminator. \citet{Sotzen_2020} analyzed HST Wide Field Camera 3 (WFC3) and  Magellan LDSS3 transmission spectra of WASP-79b, alongside photometry from TESS and Spitzer, identifying H$_2$O absorption in the infrared and attributing strong optical absorption to FeH. An independent analysis by \citet{Skaf_2020} reached similar conclusions. Recently, \citet{Rathcke_2021} used additional HST STIS observations to extend WASP-79b's transmission spectrum into the near-UV. Their analysis concluded that bound-free absorption from H$^{-}$, combined with unocculted stellar faculae, provide a better fit than FeH to the full near-UV to infrared transmission spectrum. All three of these analyses required optical absorbers -- be it FeH or H$^{-}$ -- with abundances significantly above those predicted by chemical equilibrium.    


Thermal emission observations from the dayside of WASP-79b can provide new insights into disequilibrium processes at play in its atmosphere \citep{Fortney_2021}. Photometric eclipse observations for WASP-79b were investigated previously by \citet{Garhart_2020} and \citet{Baxter_2020}, using data collected by the \textit{Spitzer Space Telescope} at wavelengths of 3.6 and 4.5~$\mu$m. From their analysis \citet{Garhart_2020} first found the ratio of planetary flux to stellar flux ($F_p/F_s$) at 3.6 and 4.5 $\mu$m as 1394 $\pm$ 88 ppm and 1783 $\pm$ 106 ppm respectively. They also determined brightness temperatures ($T_b$) at these wavelengths, finding the 3.6\,$\mu$m band to have a $T_b$ of 1959\,$\pm$\,125 K and the 4.5\,$\mu$m band to have a $T_b$ of 1948\,$\pm$\,117 K. In \citet{Baxter_2020} they utilized the flux ratio found by \citet{Garhart_2020} and through their analysis found the brightness temperatures to be 1893\,$\pm$\,49 K at 3.6\,$\mu$m and 1882\,$\pm$\,54 K at 4.5\,$\mu$m. These findings in \citet{Garhart_2020} are consistent with WASP-79b emitting as a blackbody with a effective temperature of roughly 1950 K, while \citet{Baxter_2020} is more in line with a blackbody emitting at 1890 K. Although both the \citet{Garhart_2020} and \citet{Baxter_2020} studies agree that the infrared dayside emission from WASP-79b is consistent with a blackbody (e.g. no clear signs of a thermal inversion), emission measurements at near-infrared wavelengths can provide important additional constraints on the processes shaping its dayside thermochemical structure.

In this paper, we investigate the dayside thermal emission from WASP-79b using HST's Wide Field Camera 3 G141 (1.1-1.7\,$\mu$m). These observations from HST allow us to probe a key H$_2$O feature along with other molecular features not accessible in the {\it Spitzer} observations. The WFC3 G141 range also covers a region of strong H$^{-}$ opacity, allowing recent evidence of H$^{-}$ from transmission spectra of WASP-79b \citep{Rathcke_2021} to be independently assessed. Our new emission spectrum thereby offers further insights into WASP-79b's atmospheric composition.

In what follows, we begin in Section~\ref{sec:obs} with a description of our observations and data collection. We then detail our analysis of the data, including our application of systematic corrections, in Section~\ref{sec:methods}. Our results are discussed in Section~\ref{sec:results}, which are interpreted and placed in a larger context in Section~\ref{sec:discussion}. Finally, in Section~\ref{sec:conclusion} we summarize our results and conclusions. 

\begin{figure*}[htb!]
\begin{center}
\includegraphics[scale=0.5]{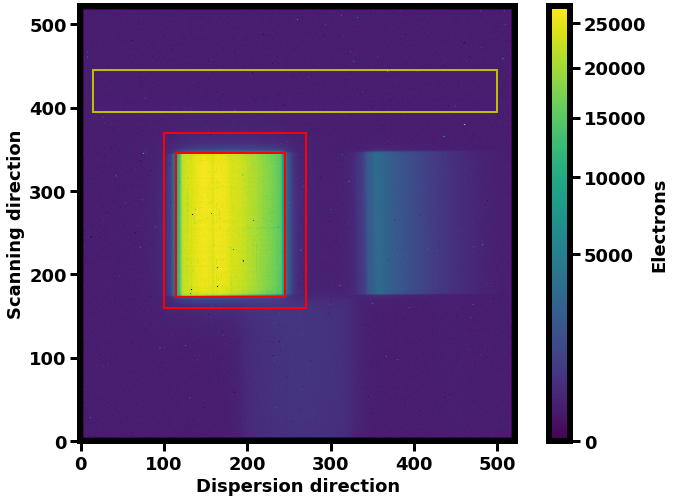}
\caption{The 2D spectrum from a frame taken during the first visit of WASP-79b before background subtraction and spectra extraction. A contaminant source is evident in the lower center of the frame and the second order spectra is seen on right side of frame. The yellow box indicates the region used for background estimation. The larger red box indicates the search area set for the first order spectra optimization. The smaller red box is the end result first order spectra box.  \label{ndr}}
\end{center}
\end{figure*}

\section{Observations} \label{sec:obs}
The observations of WASP-79b presented here, were taken as part of HST's Panchromatic Exoplanet Treasury (PanCET) program GO-14767 (PIs David Sing and Mercedes L\'{o}pez-Morales). The target was observed on November 15th 2016, using the HST WFC3 G141 grism, to acquire spectra from 1.1 to 1.7 $\mu$m. Observations were completed over five consecutive HST orbits with the first orbit having 12 exposures and the subsequent four orbits collecting 13 exposures each, resulting in a total of 64 exposures. Data were collected in forward spatial scan mode with a scan rate of $\sim$0.65 pixels/s, utilizing the 512\,$\times$\,512 subarray, in SPARS25, with 7 non-destructive reads per exposure, resulting in an exposure time of 138.38 seconds. Due to the timing of the eclipse and the HST orbit, neither the planetary ingress nor egress were captured with two orbits conducted pre-eclipse, two orbits during eclipse, and one orbit post-eclipse.

\section{Methods} \label{sec:methods}
For our analysis we used the \textit{IMA} files, that are produced from the CalWF3 pipeline which calibrates for zero-read bias and dark current subtraction. Spectra extraction was completed following the method outlined in \citet{Evans_2016}. During the five HST orbits, observations occurred prior-, during-, and post-eclipse of WASP-79b. The first orbit was discarded as is standard procedure with HST/WFC3 light curves, due to occurrence of vastly different systematics not seen in subsequent orbits \citep[e.g.][]{Deming_2013, Wakeford_2013,Sing_2016}. The first exposure of each subsequent orbit was also removed before analysis. The reason for this is that the buffer is dumped between orbits which leads to the first exposure in each orbit to contain significantly lower count levels than subsequent exposures \citep{Wakeford_2017}. Lastly, through visual inspection it was identified that the second exposure of the second orbit similarly had significantly lower count levels and was also removed. After these removals, we were left with a total of 47 exposures over four HST orbits. The time of observations were also converted from $\text{MJD}_{\text{UTC}}$ to $\text{BJD}_{\text{TDB}}$ following the prescription by Eastman \citep{Eastman_2010} utilizing barycorrpy \citep{Kanodia_2018}.

\subsection{``White" light curve}
Construction of the band integrated or ``white" light curve (WLC), first involved extracting boxes inside each exposure that corresponded to the first order spectra. To find the box edges a search region was established visually on the first non-destructive read of the first exposure that ensured neither the second order spectra nor the contaminant source shown in Figure~\ref{ndr} would be included in the search. Within the defined search area a scan was performed through each column and then row to identify where the detector counts jumped to several times the background count. The four edges were then found as the median pixel value found along each respective edge. Once the first order spectra box was defined each pixel was flat field corrected and background counts were removed. Cosmic rays were then masked using sigma-clipping for any pixel more or less than 5 standard deviations from the central value. The WLC is then produced by summing all of the pixels within this box for each exposure. From here the flux measurements were normalized by dividing each point by the mean flux of the remaining exposures.

\begin{figure*}[htb!]
\begin{center}
\includegraphics[scale=0.5]{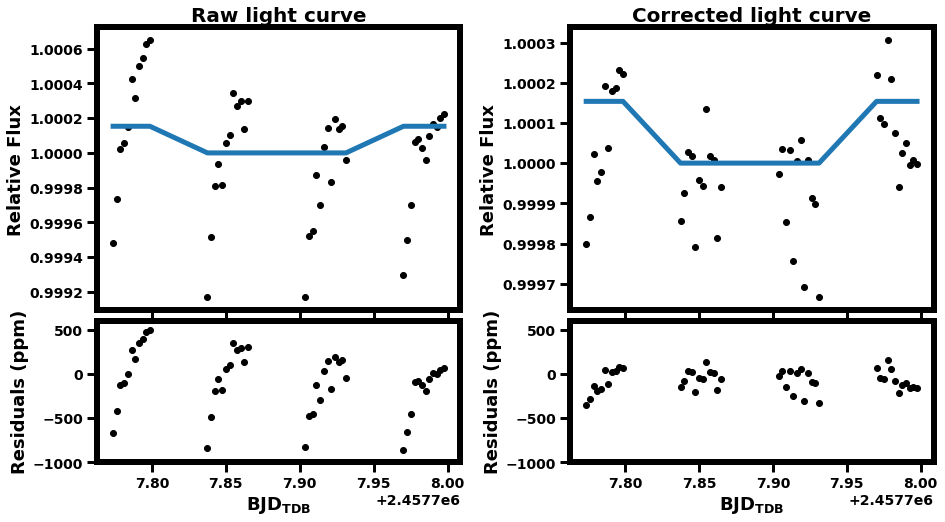}
\caption{Band-integrated ``white" light curve and residuals for the model where $t_{sec}$ is a free parameter. Top left: Raw WLC (black dots) and eclipse model (blue). The eclipse model was made using BATMAN \citep{Kreidberg_2015} with the orbital parameters from Table \ref{orb params}. Bottom left: Residuals between the raw data and model in parts per million (ppm), showing both the linear and hook systematics present in the data.  Top right: WLC after systematic corrections (black dots) with eclipse model (blue). Bottom right: Residuals between corrected data and model in ppm.\label{wlc}}
\end{center}
\end{figure*}

The raw light curves for WASP-79b (Figure~\ref{wlc}) exhibit similar instrumental systematics seen in previous WFC3 data \citep[e.g.][]{Berta_2012, Wakeford_2016, Kilpatrick_2018, Mansfield_2018}.  These systematics include a visit-long linear component and an orbit-long exponential ramp component, which are due to charge carrier trapping within the HgCdTe arrays of WFC3 \citep{Zhou_2017}. The charge carrier trapping rates are proportional to total incoming flux and so these systematics are wavelength-independent. For this reason we modeled the correction factors off the WLC. 

We fit the WLC with a model of the form:
\begin{equation}
\label{eqn:model}
    F(t) = F_mL(t)H(t)
\end{equation}
where $F(t)$ is the measured flux at time t; $F_m$ is the eclipse model \citep{Kreidberg_2015}, $L(t) = mt + b$ is the time-dependent linear component, with $m$ and $b$ as free parameters; and $H(t) = 1.0 - e^{-SP+\phi} + CP$ is the orbit-long hook component, modeled as a rising exponential where $P$ represents the HST orbital phase and $S$, $\phi$, and $C$ are free parameter \citep{Sotzen_2020}.  For the eclipse model, we consider the eclipse depth, $F_p/F_s$, and center of eclipse time, $t_{sec}$, as free parameters and fix the orbital period, planet radius, semi-major axis, orbital inclination, eccentricity, and longitude of periastron to the values determined by \citet{Sotzen_2020} as these values cannot be well constrained from a single eclipse observation. These orbital parameters are summarized in Table \ref{orb params}.

\begin{deluxetable}{cc}[htb!]
\tablecaption{Orbital parameters used in the eclipse model. Values determined by \citet{Sotzen_2020}. \label{orb params}}
\tablehead{
\colhead{Parameter} & \colhead{WASP-79b}
}
\startdata
$P_{orb}$ & 3.66239264 days \\
$R_p/R_s$ & 0.10675 \\
$a/R_s$ & 7.292 \\
$i$ & $85.929  ^{\circ}$ \\
$e$ & 0.0 \\
$\omega$ & $90  ^{\circ}$ 
\enddata
\end{deluxetable}

Due to the small amplitude of the eclipse signal and the fact that planetary ingress and egress were not captured in our observations, we considered fits to the data with $t_{sec}$ both as a fixed and free parameter. In the method where $t_{sec}$ is fixed, the fixed value was determined by taking the center of transit time and orbital period found in \citet{Sotzen_2020} and projecting it forward in time to when the eclipse was expected to occur within the observation window. For the forward projection calculation it was assumed that the eccentricity of the planet was zero as this is predicted to be common for hot Jupiters due to tidal orbital circularization \citep[e.g.][]{Dawson_2018}. The calculations were made under the assumption that the eclipse occurs half an orbital period after the transit. As shown in \citet{Kilpatrick_2017} and \citet{Cowan_2011} even under the assumption of a zero eccentricity orbit, small offsets in the center of eclipse time are measured as a result of the non-homogenous temperature structure of the planet's dayside. Additionally an offset from 0.5 orbital phase could be introduced due to the light-travel time across the system \citep[e.g.][]{Williams_2006, Dobbs_2015}. For this reason we also consider a second fit to the data with a free center of eclipse time.

In both models the free parameters and their uncertainty estimates were determined utilizing \textit{emcee's} Markov Chain Monte Carlo (MCMC) algorithm \citep{Foreman_2013}. Each run utilized 50 walkers that were initialized randomly within the parameter space such that each parameter's starting position was within one standard deviation of the initial best fit value. The initial best fit values were determined using a least-squares fit to the data. Each walker performed 10,000 steps with the first 20\% of steps removed for burn in. For the MCMC algorithm, a uniform prior with bounds of 0 and 1 was placed on both models' eclipse depth parameter to prevent exploration of nonphysical parameter space. For the model where the secondary eclipse time was a free parameter, a Gaussian prior was placed on it centered around the expected eclipse time with a width of 7.7 minutes based on eclipse timing and standard deviation from \citet{Garhart_2020}. This prior was placed to ensure the walkers only explored the parameter space consistent with the Spitzer eclipse timings from \citet{Garhart_2020}. Besides these priors all other parameters had uninformed priors. 

\subsection{Spectroscopic light curves}
\begin{figure}[htb!]
\begin{center}
\includegraphics[scale=0.5,angle=0]{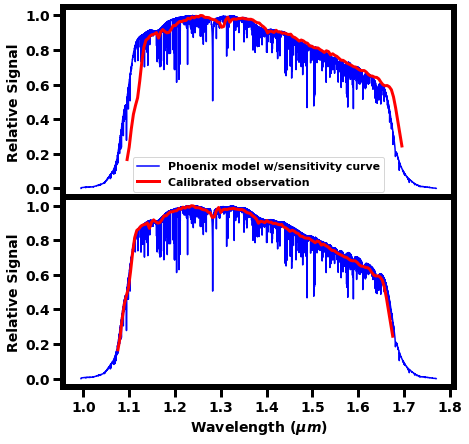}
\caption{Observed relative spectrum (red) along with the WFC3 G141 grism sensitivity curve convolved with relative Phoenix model (blue). Top: Spectrum after initial calibration following \citet{Kuntschner_2009}. Bottom: Empirically adjusted spectrum aligned with Phoenix model with sensitivity curve. \label{wlcal}}
\end{center}
\end{figure}

To construct the spectroscopic light curves, we were unable to perform the standard wavelength calibration based on the filter image obtained of the target star as it was measured with a different subarray. This resulted in offsets not accounted for in the standard wavelength calibration. Instead, we followed the procedure outlined in \citet{Wilkins_2014}. Initial calibration was completed with an estimated centroid and coefficients provided in the Space Telescope Science Institute (STScI) calibration report \citep{Kuntschner_2009}. The resulting calibrated observations were compared with the sensitivity curve provided by STScI for the G141 grism convolved with a Phoenix stellar model for WASP-79 \citep{Phoenix_2016}. As seen in the top graph of Figure \ref{wlcal} the two curves are not well aligned after initial calibration, so a shift in the calibration was made empirically using the two ends of the spectrum and the Pa-$\beta$ hydrogen line at 1.282\,$\mu m$ as guides. The offset adjusted wavelength calibration is shown in the bottom graph of Figure \ref{wlcal}.

Observations were then divided into three equal wavelength bins of width 0.20\,$\mu m$ centered at 1.18, 1.38, and 1.58\,$\mu m$ to create the spectroscopic light curves. Given the low amplitude of the expected signal, we selected three bins to sample in and out of the expected water feature $\sim$1.4~$\mu$m. The systematics within these light curves were corrected using the \textit{Divide-White} method described in \citet{Stevenson_2014}. This method assumes that the systematics are wavelength independent, and as such the spectroscopic curves are divided by the systematics from the white light curve. 

After the \textit{Divide-White} method had been performed, an observation-long wavelength dependent linear systematic was identified. Each spectroscopic light curve was individually corrected for this systematic with a first order polynomial that was unique to the wavelength bin \citep[e.g.][]{Kilpatrick_2018}. The remaining residuals were then examined for additional wavelength dependent systematics but none were identified. Similar analysis was done for each of the spectroscopic light curves as described for the white light curve using MCMC, with both $F_p/F_s$ and $t_{sec}$ set as free parameters. Corner plots for all MCMC results can be found on Zenodo\footnote{\href{https://zenodo.org/record/4662098\#.YV3rBoDMLJw}{Online supplementary figures}}.

\newpage

\subsection{Brightness Temperature and Atmospheric Forward Modeling}
Analysis on the brightness temperature of WASP-79b was done by comparing models of planet-to-star flux ratios to the two wavelength bands observed by {\it Spitzer} at 3.6 and 4.5\,$\mu$m and the three spectral channels computed with the HST/WFC3 data at 1.18, 1.38, and 1.58\,$\mu$m and the white light centered at 1.4\,$\mu$m. To calculate the models used to create the planet-to-star flux ratios, we assumed the host star to be a blackbody with a stellar temperature of 6600 K as listed in the planet's discovery paper \citep{Smalley_2012}. For the planet's emission we considered a range of brightness temperatures from 1500 K to 2000 K. For our comparison to theoretical atmospheric models we considered three types of forward models: 1D radiative-convective equilibrium; disequilibrium chemistry; and cloudy models.

In our comparison of theoretical equilibrium atmospheric models we utilized spectra generated from one-dimensional radiative-convective equilibrium atmospheric models generated specifically for WASP-79b from \citet{Goyal_2020}. When modeling the atmospheres of hot Jupiters like WASP-79b it is often assumed that their atmospheres are in chemical equilibrium due to their short radiative and chemical timescales \citep[e.g.][]{Burrows_1999, Barman_2007,Goyal_2020}. These atmospheric forward models consider four different recirculation factors (RCF = 0.25, 0.5, 0.75, 1.0), six metallicities (0.1$\times$, 1, 10, 50, 100, 200; all in solar) and six carbon-to-oxygen ratios (C/O = 0.35, 0.55, 0.7, 0.75, 1.0, 1.5). The recirculation factor describes the redistribution of incoming stellar radiation from the dayside of the planet to the nightside through means of advection from wind. A recirculation factor of 1.0 corresponds to no redistribution, whereas a RCF of 0.25 indicates 75\% of the incoming stellar energy is redistributed to the nightside through advection \citep{Goyal_2020}. Both metallicity and C/O ratio help to provide insight into a planet's formation mechanism and location. The metallicity of a planet indirectly effects the chemical composition of the atmosphere, which affects spectral features that can be observed. Based on core accretion models we'd expect gas giants like WASP-79b to have metallicities around 1--10$\times$ solar \citep[e.g.][]{Thorngren_2016, Thorngren_2019}. More specifically \citet{Thorngren_2019}, have used interior modelling to determined an upper limit on the atmospheric metallicity for WASP-79b to be 50.99$\times$ solar. The C/O ratio is also related to chemical composition of the atmosphere and is affected by planetary formation in relation to different snowlines, predominantly H$_2$O, CH$_4$, and CO \citep{Oberg_2011}. For solar, the C/O ratio is 0.56.

Since transmission observations of WASP-79b suggest that disequilibrium chemistry \citep{Sotzen_2020, Rathcke_2021} and cloud formation \citep{Stevenson_2016} might play a role in WASP-79b's atmosphere, we additionally constructed forward models to explore how these processes might shape its dayside emission spectra. Using the equilibrium chemistry models described above as a starting point, we generated spectra using enhanced abundances of FeH and H$^{-}$, species seen in the analyzes by \citet{Sotzen_2020} and \citet{Rathcke_2021} respectively, as well as VO. The inclusion of VO disequilibrium models was chosen as its affect on the eclipse depth would be similar to that of FeH at the wavelength's probed by Spitzer but begin to differentiate near 1.4\, $\mu$m where the HST/WFC3 observed. 

\def\arraystretch{1.5}
\tabletypesize{\small}
\begin{deluxetable*}{l|ccccc}[htb!]
\tablecaption{Results with $1\sigma$ uncertainties from MCMC of white light and spectroscopic models for HST/WFC3 eclipse observations of WASP-79b. \label{results}}
\tablehead{
\colhead{} & \colhead{fixed $t_{\rm sec}$} & \colhead{free $t_{\rm sec}$}\\
\colhead{} & \colhead{WLC} & \colhead{WLC} & \colhead{1.18\,$\mu$m} & \colhead{1.38\,$\mu$m} &\colhead{1.58\,$\mu$m}
}
\startdata
$F_p/F_s~ (\text{ppm})$ &
$154^{+64}_{-64}$ &
$154^{+62}_{-61}$ &
$106^{+87}_{-68}$ &
$184^{+71}_{-70}$ &
$188^{+96}_{-91}$ \\
$t_{\rm sec}$ ~ ($\text{BJD}_{\text{TDB}}-2,450,000$) & 
7707.8816 & 
$7707.8853^{+2.7e-3}_{-2.7e-3}$ &
$7707.8853^{+2.7e-3}_{-2.7e-3}$ & 
$7707.8853^{+2.7e-3}_{-2.7e-3}$ & 
$7707.8853^{+2.7e-3}_{-2.7e-3}$ \\
\enddata
\end{deluxetable*}

In order to explore the potential of clouds to shape WASP-79b's dayside emission, we leveraged the open-source exoplanet cloud formation code \texttt{Virga}\footnote{\url{https://github.com/natashabatalha/virga}} \citep[with methodology based on][]{Ackerman_2001} and atmospheric radiative transfer code \texttt{PICASO} \citep{Batalha_2019, Batalha_2020} to 
generate cloudy emission spectra. The one-dimensional radiative transfer in \texttt{PICASO} is computed taking into account the optical and scattering properties of clouds.
\texttt{Virga} chooses cloud condensate species for which the partial pressure is greater than the vapor pressure, considering vertical mixing and atmospheric chemical composition. The sedimentation efficiency ($f_{sed}$), used in \texttt{Virga} and defined in \citet{Ackerman_2001}, dictates the efficiency of the atmosphere to deposit cloud particles towards higher pressures. Higher values of sedimentation efficiency give clouds that are thinner and more depleted with larger particles, while smaller values correspond to clouds that extend higher and deeper in the atmosphere with smaller particles. 
 We compute one-dimensional emission spectra of WASP-79b by setting a constant value of $3\times10^{10}\ cm^2 s^{-1}$ for the eddy mixing coefficient based on estimates from general circulation models \citep{Kataria_2016}, which determines the vertical mixing in the atmosphere. Given the low sedimentation efficiencies expected in close-in exoplanet atmospheres \citep[e.g.][]{Morley_2015, Gao_2018}, we ran several models with different values of $f_{sed}$ ranging from 0.001 to 1.0 in order to identify simulated spectra that best matches the data. Using the thermal structure of WASP-79b predicted by the atmospheric models of \citet{Goyal_2020} assuming solar metallicity, solar C/O, and recirculation factors of 0.75 and 1.0, \ce{Al_2O_3}, Fe, and \ce{TiO_2} bearing cloud species are allowed to form.

\section{Results} \label{sec:results}
The results for the WLC, show good agreement between the two models, where $t_{sec}$ was either a fixed or free parameter. In the model where $t_{sec}$ was fixed to $2457707.8816 ~\text{BJD}_{\text{TDB}}$, $F_p/F_s$ was determined to be $154^{+64}_{-64}$~ppm. While in the model where $t_{sec}$ was a free parameter, $t_{sec}$ was found to be $2457707.8853 \pm{0.0027} ~\text{BJD}_{\text{TDB}}$ and $F_p/F_s$ was $154^{+62}_{-61}$~ppm. The $\chi^2_{red}$ for both models is 0.94.  In subsequent analysis, the model with $t_{sec}$ as a free parameter was utilized. From the spectral analysis we find no difference in $t_{sec}$ from that of the WLC in any of the three spectral channels. Table \ref{results} summarizes the secondary eclipse depths for the wavebands with their $1\sigma$ eclipse depth uncertainties.

\begin{figure*}[htb!]
\begin{center}
\includegraphics[scale=0.65]{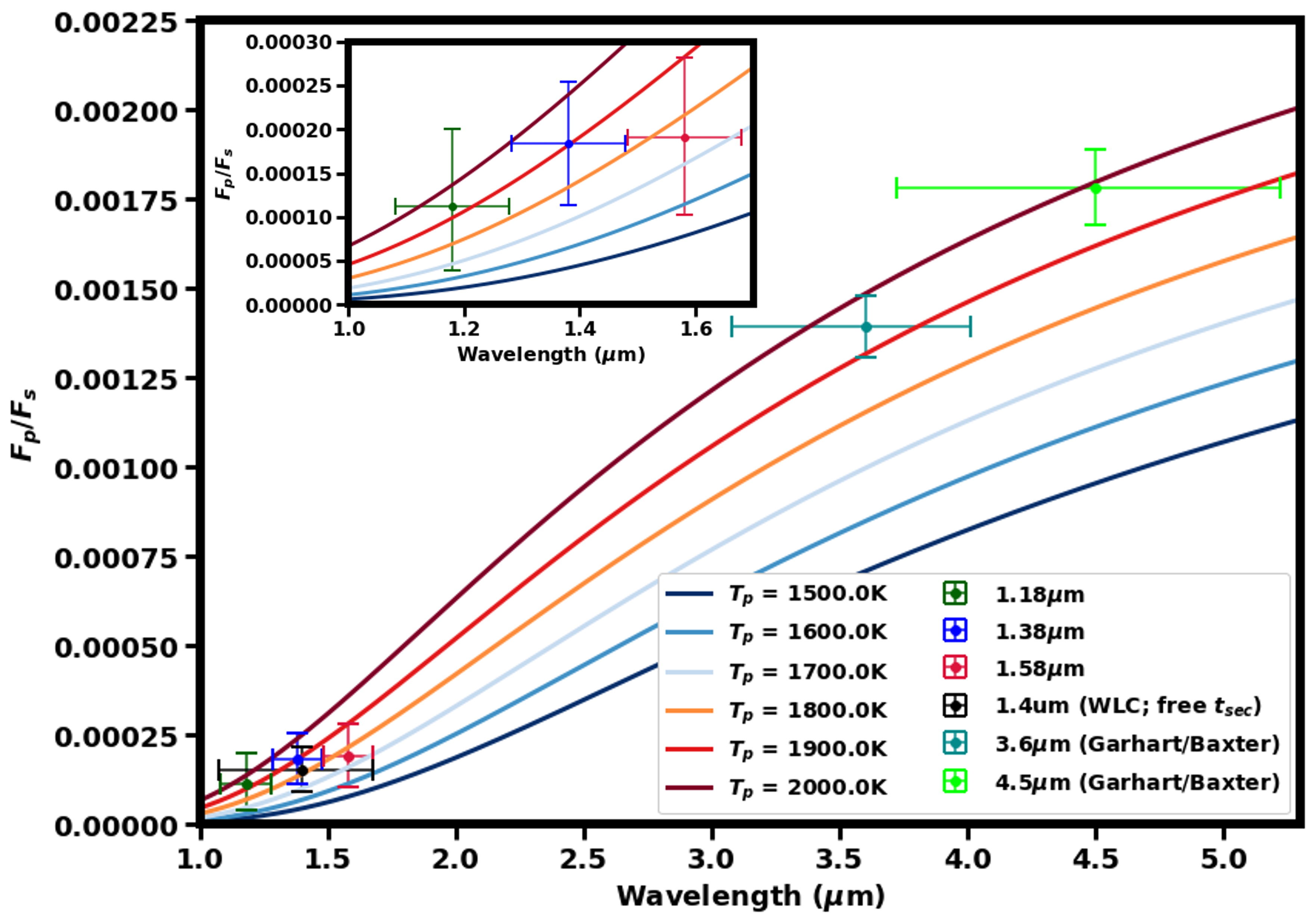}
\caption{Observed planet-to-star flux ratios/eclipse depths for WASP-79b, with observations from \citep{Garhart_2020, Baxter_2020} using {\it Spitzer} data at 3.6~$\mu$m and 4.5~$\mu$m, along with HST/WFC3 G141 data reduced herein. Plotted over the observations are models for planet-to-star flux ratios assuming a stellar blackbody with a temperature of 6600~K \citep{Smalley_2012} and a range of planetary blackbodies with temperatures ranging from 1500~K to 2000~K in 100~K increments as indicated by the figure legend. Inset highlighting just the HST/WFC3 G141 eclipse depths from this paper. Note that while the HST and {\it Spitzer} measured eclipse depths for WASP-79b are consistent with a roughly 1900~K blackbody, the HST points suggest that the WFC3 G141 bandpass is probing a slightly cooler layers of the atmosphere compared to those being probed by the longer wavelength {\it Spitzer} observations.}
\label{BB_fig}
\end{center}
\end{figure*}

We combined the estimates for $F_p/F_s$ of WASP-79b both from our analysis of the WLC at 1.4~$\mu$m and the three spectral bins at 1.18, 1.38, and 1.58~$\mu$m with the results from \citet{Garhart_2020} and \citet{Baxter_2020} at 3.6 and 4.5~$\mu$m and compared them with theoretical predictions to better understand the physical properties of WASP-79b's atmosphere. Our first comparison shown in Figure \ref{BB_fig}, compared the calculated $F_p/F_s$ points to blackbody curves assuming a range of temperatures for WASP-79b. From this comparison, WASP-79b appears to fit well as a blackbody with an effective temperature of around 1900~K with the exception of the 1.58~$\mu$m observation which looks to be probing slightly lower temperatures closer to 1800~K. 

\begin{figure}[htb!]
\begin{center}
\includegraphics[scale=0.37]{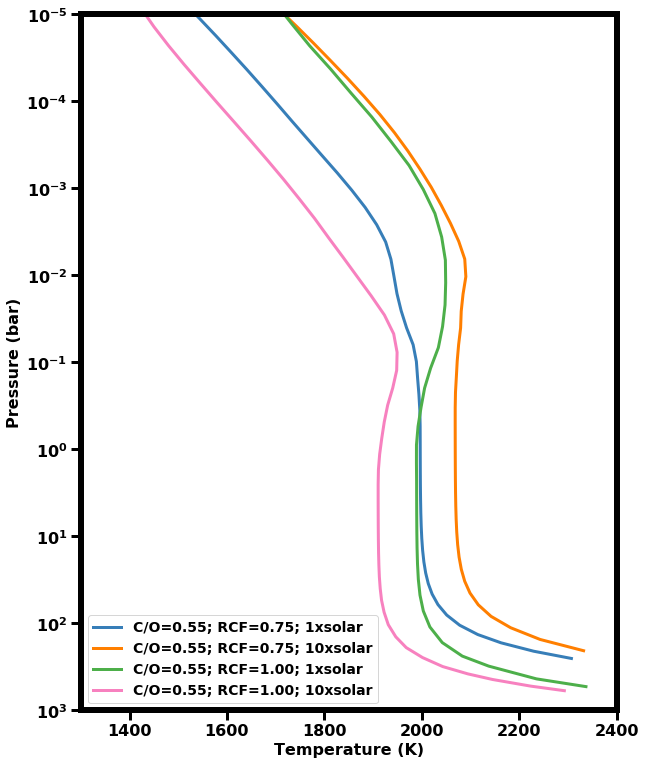}
\caption{Pressure-temperature profiles used in the calculations of the emission models from \citep{Goyal_2020} that are highlighted in Figures \ref{Eq_model}, \ref{Diseq_Model}, and \ref{Eq_model_VIRGA}.
\label{PT_fig}}
\end{center}
\end{figure}

Next we compared our data to a library of possible atmospheres of WASP-79b developed by \citet{Goyal_2020} through a chi-square analysis. Figures \ref{PT_fig} and \ref{Eq_model} shows the pressure-temperature profiles and emission spectra for a subset of the equilibrium chemistry atmospheric models we considered in this study. Note that in considering disequilibrium chemistry and cloud formation in WASP-79b's atmosphere (Figures \ref{Diseq_Model} and \ref{Eq_model_VIRGA} ) we assume that the underlying thermal structure of the atmospheres is the same as in the relevant equilibrium chemistry atmospheric models.

Results from this comparison are summarized in Table \ref{chisq}, which shows the seven best fit models to the observed data, listing the values of the three free parameters explored: recirculation factor, metallicity, and carbon-to-oxygen ratio and the resulting reduced chi-square for each model. From this comparison, we find that models with RCF of 0.75 and metallicity of 1$\times$ solar are preferred with solar or slightly above solar C/O ratio.

\def\arraystretch{1.25}
\tabletypesize{\small}
\begin{deluxetable}{cccc}[htb!]
\tablecaption{List of best fit equilibrium chemistry models from \citep{Goyal_2020} with their calculated reduced chi-square. 
\label{chisq}}
\tablehead{
\colhead{RCF} & \colhead{log(Z)} & \colhead{C/O ratio} & \colhead{$\chi^2_r$}
}
\startdata
$0.75$ &
$0$ &
$0.7$ &
$1.04$ \\
$0.75$ &
$0$ &
$0.55$ &
$1.09$ \\
$0.75$ &
$0$ &
$0.35$ &
$1.54$ \\
$0.75$ &
$1.0$ &
$0.75$ &
$2.14$ \\
$0.75$ &
$2.0$ &
$0.75$ &
$2.2$ \\
$0.75$ &
$1.0$ &
$0.55$ &
$2.29$ \\
$1.0$ &
$0$ &
$1.5$ &
$2.88$ \\
$1.0$ &
$0$ &
$0.55$ &
$10.21$ \\
$1.0$ &
$1.0$ &
$0.55$ &
$7.49$
\enddata
\end{deluxetable}

We then compared our observational data with model spectra assuming disequilibrium abundances for FeH, VO, and H$^-$ (Figure \ref{Diseq_Model}) and find that several of these disequilibrium models can explain the observations just as well as the equilibrium models shown in Figure \ref{Eq_model}. Finally, we compared our observational data to model spectra that assume a chemical composition and thermal structure identical to the chemical equilibrium atmospheric models, but where opacity due to cloud formation is considered and find that cloudy model spectra ($f_{sed}=0.002$) can also adequately explain the observations (Figure \ref{Eq_model_VIRGA}). 

\section{Discussion} \label{sec:discussion}
\subsection{Secondary eclipse timing}
We chose to investigate our eclipse signal under the assumption of both fixed and free center of eclipse timing due to the weakness of this signal in the HST/WFC3 G141 data. As previously noted we expect WASP-79b to have an eccentricity near zero due to tidal circularization allowing us to fix $t_{sec}$ based upon previous calculations of transits and their periodicity. However, although we do not expect a large eccentricity, it is not a well constraint parameter for WASP-79b. There are also other effects that can shift the observed $t_{sec}$, such as, an off-center hot spot on the planet or the light-travel time across the system \citep[e.g.][]{Williams_2006, Dobbs_2015}. 
When we compare the center of eclipse times between the two methods we find that the unconstrained secondary eclipse time occurs about 5 minutes later, suggesting a possible offset from the 0.5 orbital phase. The offset is within $2\sigma$ of the fixed $t_{sec}$ value however, so additional observations will be needed to confirm with greater confidence the evidence of this offset. 

\subsection{Blackbody models}
Proceeding with eclipse depths measured under the condition where $t_{sec}$ is allowed to be a free parameter, we then explored the possible insights into WASP-79b's atmosphere that could be gleaned from the combination of 1.1-1.7~$\mu$m eclipse depths measured in this study and the 3.6 and 4.5~$\mu$m eclipse depths presented in \citet{Garhart_2020} and \citet{Baxter_2020}. As shown in Figure \ref{BB_fig}, the HST and Spitzer eclipse depths are all consistent with a blackbody between 1800~K and 2000~K. In translating the WASP-79b {\it Spitzer} eclipse depths into brightness temperature \citet{Garhart_2020} suggest a brightness temperature around 1950~K while \citet{Baxter_2020} suggest a slightly lower brightness temperature around 1890~K. The HST points are in good agreement with both {\it Spitzer} derived brightness temperatures, however, the WLC point at 1.4~$\mu$m and the spectral point at 1.58~$\mu$m do suggest the possibility of probing a cooler temperature with HST, which may suggest a weak thermal inversion in the pressure-temperature profile of WASP-79b's dayside shown in Figure \ref{PT_fig}. 

\begin{figure*}[htb!]
\begin{center}
\includegraphics[scale=0.625]{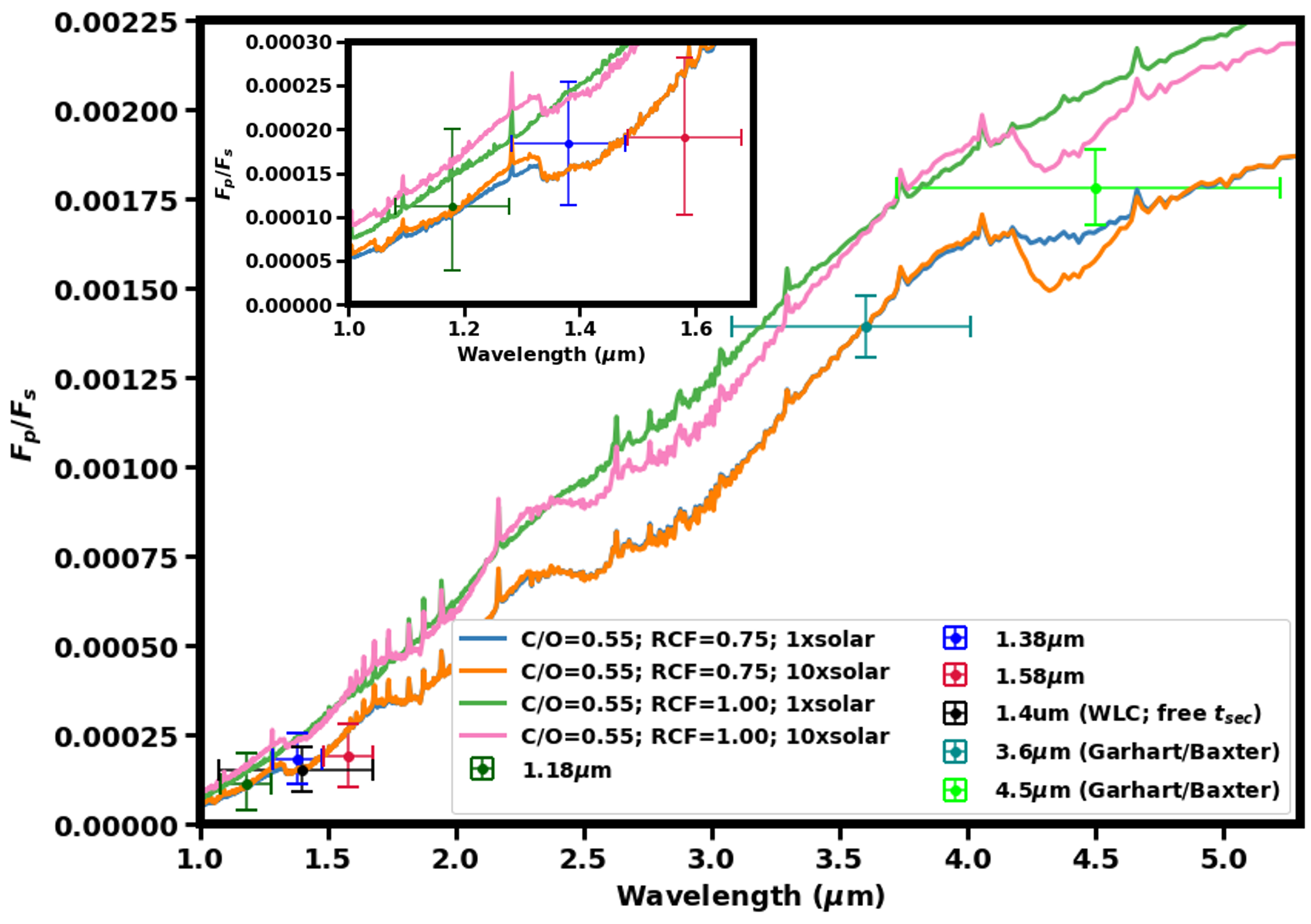}
\caption{Observed planet-to-star flux ratios/eclipse depths for WASP-79b, with observations from \citep{Garhart_2020, Baxter_2020} using {\it Spitzer} data at 3.6~$\mu$m and 4.5~$\mu$m, along with HST/WFC3 G141 data reduced herein. Plotted over the observations are four possible equilibrium atmospheric models from \citet{Goyal_2020}. All four models have the same C/O ratio of 0.55 but vary with recirculation factors of 0.75 and 1.00 and metallicities of 1$\times$ and 10$\times$ solar. Here it is clear that the observations favor atmospheric models with a RCF = 0.75, suggesting $\sim 25\%$ of the incoming stellar energy is transported from the day to night side of the planet. Of those models the observation at 4.5~$\mu$m helps break the degeneracy in metallicity favoring the 1$\times$ solar model.
\label{Eq_model}}
\end{center}
\end{figure*}

\subsection{Forward atmospheric models}
Although HST and {\it Spitzer} dayside emission measurements for WASP-79b are roughly consistent with an isothermal blackbody spectrum as has been seen for many hot Jupiters \citep[e.g][]{Mansfield_2018, Nikolov_2018}, it is still useful to explore other atmospheric scenarios that would also be consistent with WASP-79b's measured dayside emission. As discussed previously, given the precision and phase coverage of our WASP-79b HST/WFC3 G141 observations we could not constrain the presence of a hot-spot offset from the eclipse timing alone, which would be indicative of the efficiency of heat transport in the planet's atmosphere. However, as shown in Table \ref{chisq} and illustrated in Figure~\ref{Eq_model}, in comparing WASP-79b's measured dayside emission with predictions from the grid of models from \citet{Goyal_2020}, it is clear that atmospheric models where some ($\sim 25\%$) of the incoming stellar energy is transported from the day to night side of the planet provide a better match to WASP-79b's observed flux.

Assuming some day to night energy transport (RCF=0.75), we also note that higher metallicity (e.g. 10$\times$~solar) equilibrium chemistry atmospheric models from the \citet{Goyal_2020} grid tend to underpredict the flux from WASP-79b at 4.5~$\mu$m, which is due to the increased abundance of atmospheric species such at CO and CO$_2$ that absorb strongly in the {\it Spitzer} 4.5~$\mu$m bandpass. This best fit of solar metallicity is consistent with predictions from \citet{Thorngren_2019} which uses interior modeling to place an upper limit on the metallicity of WASP-79b to $50.99\times$ solar and a bulk metallicity of $28.5 \pm{12.16}$. As highlighted in \citet{Thorngren_2019} a planet's bulk metallicity serves as an upper limit on the measured atmospheric metallicity. Since our best fit of solar metallicity is on the lower end of the metallicity range predicted by \citet{Thorngren_2019} this could indicated that "metals" are not well mixed between the interior and atmosphere of WASP-79b.
Overall, in the HST/WFC3 G141 bandpass, the measured flux from WASP-79b's dayside is well matched by the equilibrium chemistry models presented in Figure~\ref{Eq_model}, but we note that our 1.58~$\mu$m channel measurement lies about 1$\sigma$ below the equilibrium chemistry model that well matches the 1.18~$\mu$m and 1.38~$\mu$m channel emission measurements.

\begin{figure*}[htb!]
\begin{center}
\includegraphics[scale=0.625]{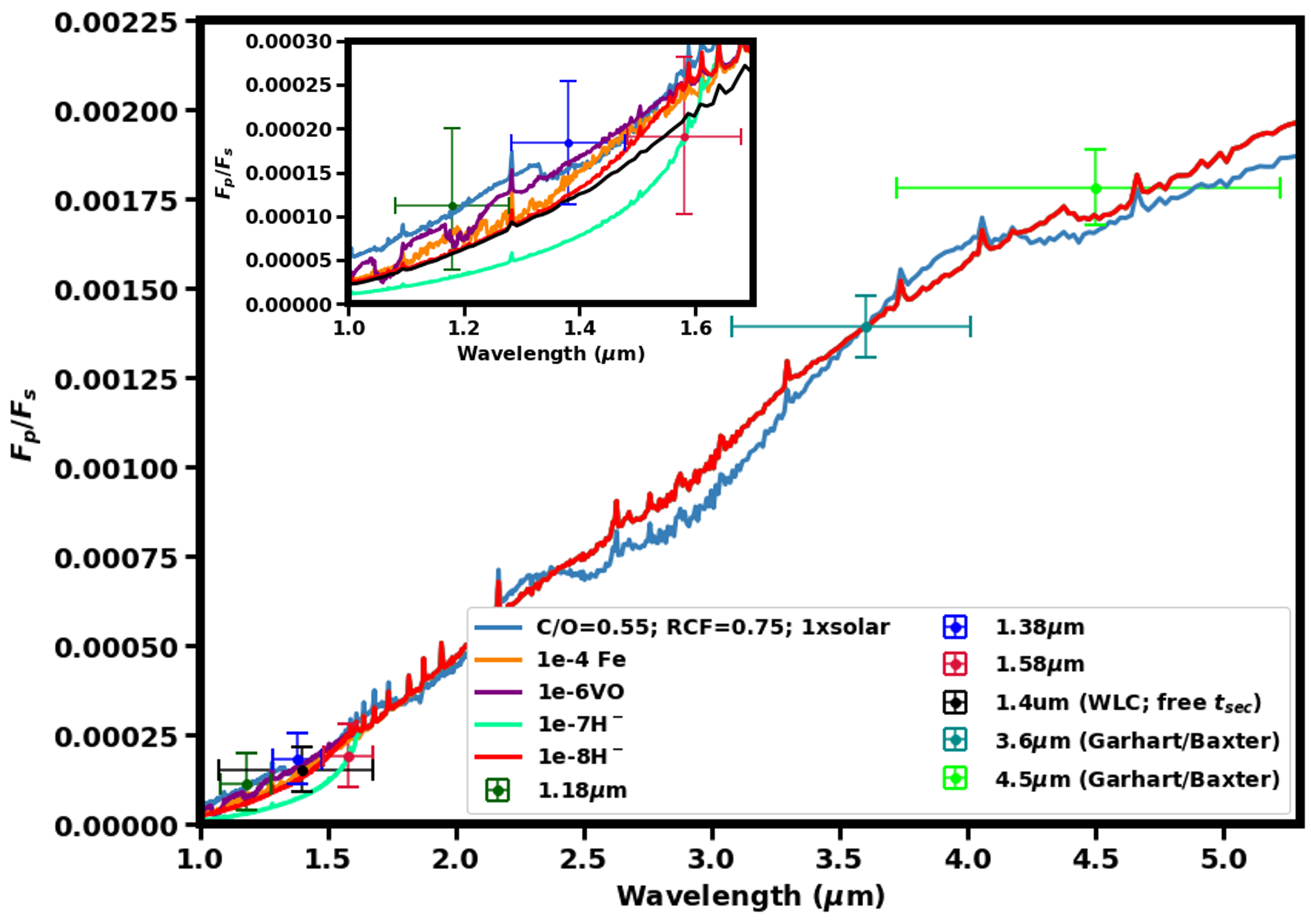}
\caption{Observed planet-to-star flux ratios/eclipse depths for WASP-79b, with observations from \citep{Garhart_2020, Baxter_2020} using {\it Spitzer} data at 3.6~$\mu$m and 4.5~$\mu$m, along with HST/WFC3 G141 data reduced herein. Plotted over the observations are possible disequilibrium atmospheric models from \citet{Goyal_2020}, that include enhanced abundances of FeH, VO, and H$^-$. There is little differentiation in the models at the {\it Spitzer} wavelengths (3.6 and 4.5~$\mu$m), however at the HST wavelengths (1.1-1.7~$\mu$m) these models become distinguishable from one another and in particular the 1.58~$\mu$m channel measurement is suggestive of a substantially increased abundance of H$^-$ in WASP-79b's dayside atmosphere with an abundance of $\sim$10$^{-7}$ compared to the $\sim$10$^{-11}$ abundance expected from equilibrium predictions.
\label{Diseq_Model}}
\end{center}
\end{figure*}

Given the atmospheric properties measured along WASP-79b's limbs via transmission spectroscopy \citep{Sotzen_2020, Rathcke_2021}, it seems possible that disequilibrium chemistry processes may be shaping WASP-79b's dayside emission as well. In Figure~\ref{Diseq_Model} we explored how the emission spectrum of WASP-79b would deviate from that of the preferred equilibrium chemistry atmospheric model from \citet{Goyal_2020} if disequilibrium abundances of key atmospheric species (FeH, VO, H$^-$) in line with transmission measurements from \citet{Sotzen_2020, Rathcke_2021} were assumed. These chemical species (FeH, VO, H$^-$) absorb more strongly at visible and near-infrared wavelengths, so it is not surprising that the {\it Spitzer} emission measurements are not sensitive to changes in the abundance of FeH, VO, and H$^-$ in WASP-79b's dayside atmosphere. In the HST/WFC3~G141 bandpass, WASP-79b's emission spectrum can be significantly affected by enhancement of FeH, VO, and H$^-$ in the planet's dayside atmosphere. Our 1.58~$\mu$m channel measurement is suggestive of a substantially increased abundance ($\sim$10$^{-7}$ vs $\sim$10$^{-11}$ from equilibrium predictions) of H$^-$ in WASP-79b's dayside atmosphere. Given that \citet{Rathcke_2021} measured H$^{-}$ abundance of 10$^{-8}$ along WASP-79b's limb, it is possible that H$^{-}$ is even more substantially enhanced in WASP-79b's dayside atmosphere. We also note that theoretical spectra with enhanced abundances of FeH, as seen in WASP-79b in transmission spectra from \citet{Sotzen_2020}, and VO are also consistent with WASP-79b's measured dayside emission. However, in all cases more precise and higher resolution spectra would be needed to substantiate claims of disequilbrium chemical abundances in WASP-79b's dayside atmosphere.


\begin{figure*}[htb!]
\begin{center}
\includegraphics[scale=0.625]{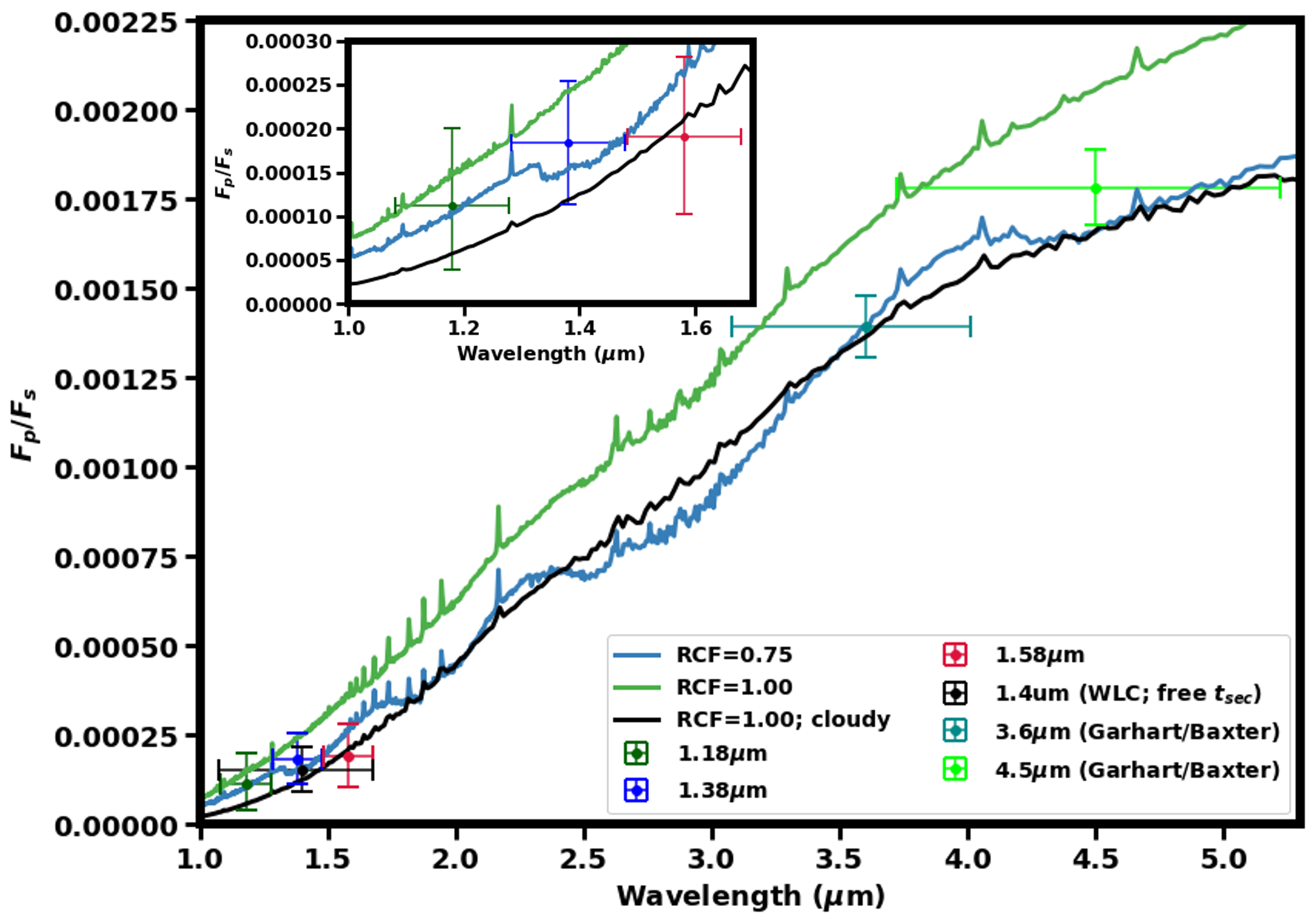}
\caption{Observed planet-to-star flux ratios/eclipse depths for WASP-79b, with observations from \citet{Garhart_2020} and \citet{Baxter_2020} using {\it Spitzer} data at 3.6\,$\mu$m and 4.5\,$\mu$m, along with HST/WFC3 G141 data reduced herein. Plotted over the observations are three equilibrium atmospheric models, all with C/O ratios of 0.55 and solar metalicities. The blue and green models are cloud-free with RCFs of 0.75 and 1.00 respectively from \citet{Goyal_2020}, while the black model has an RCF of 1.00 and includes clouds. Note that the model with a RCF of 1.00 that includes clouds closely resembles the cloud-free model with a RCF of 0.75. 
\label{Eq_model_VIRGA}}
\end{center}
\end{figure*}

Although the transmission spectra of WASP-79b presented in \citet{Sotzen_2020} and \citet{Rathcke_2021} do not suggest that clouds play a major role WASP-79b's atmosphere along its limbs, it is still possible for high-temperature condensate clouds \citep{Wakeford_2017_clouds} to form on WASP-79b's dayside and influence its emission spectrum. It is important to note that previous studies have highlighted that cloud formation can sometimes cause over or underestimates of recirculation efficiencies for hot Jupiters when comparison between observations and cloud-free atmospheric models are made \citep[e.g.][]{Parmentier_2021}. In Figure~\ref{Eq_model_VIRGA} we highlight that a theoretical emission spectrum derived from an atmospheric model that assumes inefficient day-to-night circulation (RCF=1.00) but also accounts for the possibility of clouds formation can mimic the spectrum derived from a cloud-free atmospheric model with more efficient day-to-night circulation (RCF=0.75). The presence of clouds in WASP-79b's infrared photosphere will result in an increased optical depth that will naturally cause observations to be dominated by cooler lower opacity regions in the upper atmosphere. Although a cloudy dayside may be present in WASP-79b's atmosphere, we note that the vertical mixing ($3\times10^{10}\ cm^2 s^{-1}$) and sedimentation efficiencies ($f_{sed}=0.002$) required to produce clouds with sufficient vertical extent and opacity represent extreme mixing in the atmosphere. More detailed three-dimensional models would be needed to explore the role both horizontal transport and vertical mixing may play in forming clouds in WASP-79b's dayside atmosphere.  

We chose not to conduct retrievals and limited our analysis to a comparison of reduced chi-square values for the forward models created by \citep{Goyal_2020} due to the limited number of observations. Given five data points the information content in the spectrum is not sufficiently high to justify even the simplest retrieval which includes five parameters for the pressure-temperature profile \citep[e.g.][]{Line_2013} and two chemical equilibrium parameters, recirculation factor, metallicity and C/O ratio. As such the simplest retrieval would require more free parameters than we currently have data points. Statistics like reduced chi square for the retrieval would be negative and therefore the results would not be statistically meaningful. A retrieval would also be difficult to constrain given the lack of spectral features currently observed in the WASP-79b spectrum.

A complementary analysis of a HST WFC3 G141 hot Jupiter emission spectrum was conducted by \citet{Nikolov_2018} for HAT-P-32b, a planet with similar properties to WASP-79b, including size ($R_p = 1.789 \pm{0.025}~ R_J,~ M_p = 0.860 \pm{0.164}~ M_J$), temperature ($T_{eq} = 1786 \pm{26}~ K$), and stellar host (F-type). Unlike WASP-79b's emission however, measurements for HAT-P-32b do show a strong spectral signature at the WFC3 G141 wavelengths. HAT-P-32b is also more bloated than WASP-79b and orbits closer to its slightly cooler host star. All of these conditions provide HAT-P-32b a deeper eclipse depth providing a stronger signal in its measurements compared to WASP-79b. For these reasons \citet{Nikolov_2018} were able to increase the number of spectral channel bins used in their WFC3 G141 data, allowing for retrievals to be completed. Their findings however, for HAT-P-32b closely match ours for WASP-79b, in that they found its spectrum can be explained by either a blackbody spectrum ($T_p = 1995 \pm{17}~K$) or with a spectrum with a modest thermal inversion. They similarly were unable to break the degeneracy within HAT-P-32b's atmosphere suggesting it could be a clear atmosphere with an absorber like VO, a cloud deck, or some combination of the two.
\newpage
\section{Conclusion} \label{sec:conclusion}

In this study we analyzed a secondary eclipse of WASP-79b as observed by HST/WFC3 G141 in the 1.1-1.7~$\mu$m band. We present eclipse depth for WASP-79b using two methods, one in which  the time of secondary eclipse is fixed based off the epoch and period found by \citet{Sotzen_2020} and one in which this parameter was free. The choice to consider a fixed secondary eclipse time was made because the weak signal-to-noise of the data reduced our confidence in the best fit to accurately capture $t_{sec}$. Both methods were found to be in good agreement with the eclipse depth (154~ppm) and time of secondary eclipse, however, the method in which $t_{sec}$ was a free parameter did find a slightly later than predicted eclipse time, suggesting a possible offset from an orbital phase of 0.5 by about 5 minutes. 

We then combined our HST/WFC3 G141 emission measurements with the {\it Spitzer} emission measurements from \citet{Garhart_2020} and \citet{Baxter_2020} and compared against a range of blackbody curves and a variety of possible atmospheric models based on the grid of WASP-79b models from \citet{Goyal_2020}. From this comparison we found WASP-79b closely fits a blackbody with an effective temperature around 1900~K which is in good agreement with both the results of from \citet{Garhart_2020} and \citet{Baxter_2020}. Our atmospheric modeling analysis found the best fit equilibrium atmospheric model suggests a solar metallicity, C/O ratio of 0.55, and a recirculation factor of 0.75. There are, however, several disequilibrium chemistry and cloudy atmospheric models investigated herein that can just as well explain the dayside emission of WASP-79b given current observational precision. These non-equilibrium models differentiate from one another most substantially in the near-infrared {\it Hubble} wavelengths, which could allow us to break model degeneracies. Therefore, to better understand the atmosphere of WASP-79b, further observations need to be completed in the near-infrared in order to improve the signal-to-noise, allowing us to better constrain $t_{sec}$ and $F_p/F_s$ and break degeneracies between various atmospheric model scenarios. Even with just two additional eclipse observations with HST, we would be able to increase the precision to less than 40 ppm providing a stronger secondary eclipse detection ($> 4\sigma$), and begin to differentiate between some of the atmospheric models shown in Figure \ref{Diseq_Model}. Future emission observations with the JWST would also certainly help to better constrain the properties of WASP-79b's dayside atmosphere and further refine atmospheric theories in the important planetary phase space it inhabits. 

\acknowledgments

Support for program GO-14767 was provided by NASA through a grant from the Space Telescope Science Institute (STScI), which is operated by the Association of Universities for Research in Astronomy, Inc., under NASA contract NAS 5-26555. 

GB acknowledges support from CHEOPS ASI-INAF agreement n. 2019-29-HH.0.

This work has been carried out in the frame of the National Centre for Competence in Research PlanetS supported by the Swiss National Science Foundation (SNSF). This project has received funding from the European Research Council (ERC) under the European Union's Horizon 2020 research and innovation programme (project {\sc Spice Dune}, grant agreement No 947634). The authors would like to acknowledge the anonymous referee for their useful comments.


\facility{HST (WFC3)}

\software{astropy \citep{2013A&A...558A..33A},  
          BATMAN \citep{Kreidberg_2015}, 
          SciPy \citep{2020SciPy-NMeth},
          barycorrpy \citep{Kanodia_2018}
         }
\newpage




\bibliography{WASP79b}{}
\bibliographystyle{aasjournal}



\end{document}